\documentclass[final,3p,times,twocolumn]{elsarticle}

\usepackage{amssymb}

\pdfoutput=1
\usepackage[table,xcdraw]{xcolor}
\usepackage{graphicx}
\usepackage{epstopdf}
\usepackage{mathrsfs}
\usepackage{amssymb}
\usepackage{verbatim}
\usepackage{color}
\usepackage{hhline}
\usepackage{stackrel}
\usepackage[pdfencoding=auto]{hyperref}
\usepackage{array,multirow}
\usepackage{colortbl}
\usepackage{hhline}
\usepackage{lipsum, babel}
\usepackage{xcolor}
\usepackage{lipsum}
\usepackage{amsmath}
\usepackage{multirow}
\usepackage[table,xcdraw]{xcolor}

\allowdisplaybreaks

\DeclareMathOperator*{\argmin}{arg\,min}
\DeclareMathOperator*{\Tr}{Tr}

\newcommand{\beq}{\begin{equation}}
\newcommand{\eeq}{\end{equation}}

\journal{Physics Letters B}

\biboptions{sort&compress}

\begin{document}

\begin{frontmatter}

\title{Identifying the Group-Theoretic Structure of Machine-Learned Symmetries}

\author[UF]{Roy T.~Forestano\fnref{contribution}} 
\author[UF]{Konstantin T.~Matchev\fnref{contribution}}
\author[UF]{Katia Matcheva\fnref{contribution}}
\author[UF]{Alexander Roman\fnref{contribution}}
\author[UF]{Eyup B.~Unlu\fnref{contribution}}
\author[UF]{Sarunas~Verner\fnref{contribution}}

\fntext[contribution]{All authors share equal contributions to this paper.}
\affiliation[UF]{organization={Institute for Fundamental Theory, Physics Department, University of Florida},
            city={Gainesville},
            state={FL},
            postcode={32611}, 
            country={USA}}

\begin{abstract}
Deep learning was recently successfully used in deriving symmetry transformations that preserve important physics quantities. Being completely agnostic, these techniques postpone the identification of the discovered symmetries to a later stage. In this letter we propose  methods for examining and identifying the group-theoretic structure of such machine-learned symmetries. We design loss functions which 
probe the subalgebra structure either during the deep learning stage of symmetry discovery or
in a subsequent post-processing stage.
We illustrate the new methods with examples from the U(n) Lie group family, obtaining the respective subalgebra decompositions. As an application to particle physics, we demonstrate the identification of the residual symmetries after the spontaneous breaking of non-Abelian gauge symmetries like SU(3) and SU(5) which are commonly used in model building.
\end{abstract}

\end{frontmatter}

\section{Introduction}
\label{sec:introduction}

Investigations of fundamental symmetries and the possible mechanisms for their violations in Nature are at the forefront of modern theoretical physics research \cite{Gross1996}. The ideas of supersymmetry \cite{Ramond:1971gb,Golfand:1971iw,Volkov:1972jx,Neveu:1971rx} and a grand unified theory (GUT) \cite{Georgi:1974sy,Pati:1974yy} represent attractive possibilities for physics beyond the standard model (SM), and have stimulated significant model-building and phenomenology efforts in the past \cite{Peskin:2008nw,Feng:2013pwa,Matchev:2019sqa,Croon:2019kpe}. The use of artificial intelligence for studying such symmetry paradigms is a tantalizing possibility which recently has been attracting a lot of interest. The initial focus was on symmetry discovery in data collected in specific physical systems, e.g., planetary systems, electrodynamics, etc.~\cite{Iten1807.10300,Wetzel:2020jan,Liu:2020omw,Dillon:2021gag}. Subsequent studies shifted to the discovery of symmetries in purely theoretical constructs as well \cite{Krippendorf:2020gny,Barenboim:2021vzh,Liu:2021azq,Desai:2021wbb,Craven:2021ems,Moskalev2210.04345,Forestano:2023fpj,Roman:2023ypv,Forestano:2023qcy,Forestano:2023ijh}. In either case, the natural language for discussing such sets of symmetries is group theory. It was shown that through a suitable choice of a loss function, it is possible to find a closed orthonormal set of symmetry generators that form a Lie algebra \cite{Forestano:2023fpj,Roman:2023ypv,Forestano:2023qcy,Forestano:2023ijh}. The natural follow-up questions to ask are: What kind of Lie algebra has been found? What is its rank? Is it semi-simple? Can it be decomposed into a direct sum of sub-algebras and how? 

The issue of subgroups and their respective subalgebras is a central one in the discussion of spontaneous symmetry breaking, whereby the symmetry of the full Lie group is reduced to that of one of its subgroups. Textbook examples from particle physics include the breaking of the electroweak $SU(2)\times U(1)$ gauge symmetry in the Standard Model (SM) to the $U(1)$ of electromagnetism, as well as various GUT breaking scenarios to the SM itself. 

The main focus of this letter is on investigating the group-theoretic structure of a machine-learned set of symmetries. The object of interest will be the set $\mathfrak s$ of $N_{\mathfrak s}$ symmetry generators $\mathbb J_\alpha$, $\alpha = 1,2,\ldots, N_{\mathfrak s}$, which can be found numerically following the recent methods in \cite{Moskalev2210.04345,Forestano:2023fpj,Roman:2023ypv,Forestano:2023qcy,Forestano:2023ijh}. The specific system exhibiting these symmetries is of no particular significance --- it could be a numerical dataset, or a theory model. Our goal will be to identify the subalgebra structure of the set $\mathfrak s$ by addressing the following questions.

\begin{itemize}
\item {\bf Allowed subalgebras.} Does the set $\mathfrak s$ contain valid subalgebras $\mathfrak h \subset \mathfrak s$, 
with $N_{\mathfrak h}<N_{\mathfrak s}$ generators? If so, what are all the possible integer values of  $N_{\mathfrak h}$?
\item {\bf Cartan subalgebra.}  What is the rank of $\mathfrak s$, i.e., what is the dimension of the {\em maximal} abelian subalgebra $\mathfrak h_c$?
\item {\bf Composition series.} Can the full symmetry algebra $\mathfrak s$ be represented as a direct sum of $h$ simple algebras as $\mathfrak h_1 \oplus \mathfrak h_2 \oplus \ldots \oplus \mathfrak h_h$, for some value of $h$?
\end{itemize}

The paper is organized as follows. In Section~\ref{sec:method} we develop the general formalism for addressing those questions. In Section~\ref{sec:SU4example} the technique is illustrated with the example of the $u(4)$ algebra. Particle physics applications are considered in Section~\ref{sec:models}, where we apply the method to gauge models exhibiting spontaneous symmetry breaking and identify the residual symmetries. Section~\ref{sec:conclusions} contains our conclusions. \ref{sec:so5} provides useful background on the $SO(5)$ subgroup of $U(4)$.

\section{Identifying subalgebra structures}
\label{sec:method}

A symmetry transformation generally acts on an arbitrary $n$-dimensional vector ${\mathbf x}\equiv \{x^{(1)}, x^{(2)},\ldots, x^{(n)}  \}$, where ${\mathbf x}\in \mathbb R^n$ for $O(n)$ groups or ${\mathbf x}\in \mathbb C^n$ for $U(n)$. In our procedure, a group transformation on a real space $\mathbb{R}^n$ or a complex space $\mathbb{C}^n$ is represented by a matrix operation acting on a set of $m$ points $\left\{\mathbf{x}\right\} \equiv \left\{\mathbf{x}_1,\mathbf{x}_2,\ldots,\mathbf{x}_m\right\}$ sampled from a finite domain. The choice of sampling distribution, domain size, and location is inconsequential. For definiteness, we use a standard normal distribution with a sample size of $m=300$.

A symmetry implies a conservation law, i.e., the invariance 
\begin{equation}
\varphi(\mathbf{x}'_i) 
= \varphi(\mathbf{x}_i), \quad \forall i \,= \,1,2,\ldots,m \, ,
\end{equation}
of an oracle function $\varphi(\mathbf{x})$ with respect to an infinitesimal transformation $\delta \mathbf f$
\begin{equation}
\mathbf{x} 
\stackrel{\delta \mathbf f}{\longrightarrow} \mathbf{x}' 
= \mathbf{x} + \delta \mathbf{x} 
\equiv \left(\mathbb{I} + \varepsilon \, {\mathbb J} \, \right) \mathbf{x},
\label{eq:xprime}
\end{equation}
where $\varepsilon$ is an infinitesimal parameter, ${\mathbb J}$ is an $n\times n$ matrix representing the learned symmetry generator, and $\mathbb{I}$ is the identity matrix. Following the notation of \cite{Forestano:2023fpj,Roman:2023ypv,Forestano:2023qcy,Forestano:2023ijh}, in what follows we distinguish between the matrix $\mathbb G$ representing a potential generator during the training process and the final learned generator $\mathbb J$ given by 
\beq
\mathbb{J} \; \equiv \;  \argmin_{\mathbb G}
\Bigl(L({\mathbb G}, \{\mathbf x\}) \Bigr) \, ,
\eeq
where $L$ is the loss function, written in terms of $\mathbb G$.

In Refs.~\cite{Moskalev2210.04345,Forestano:2023fpj,Roman:2023ypv,Forestano:2023qcy,Forestano:2023ijh} it was demonstrated that with a suitable choice of the loss function $L$, one can obtain the full algebra $\mathfrak s \equiv \{J\}$ of orthonormal generators for the $SO(n)$, $SU(n)$ and exceptional groups $G_2$, $F_4$ and $E_6$. We shall not repeat the details of the training procedure here, and simply refer the interested reader to \cite{Moskalev2210.04345,Forestano:2023fpj,Roman:2023ypv,Forestano:2023qcy,Forestano:2023ijh}. 
For our purposes it is worth noting that the methods of \cite{Moskalev2210.04345,Forestano:2023fpj,Roman:2023ypv,Forestano:2023qcy,Forestano:2023ijh} fall into two major categories, depending on whether the generators $\mathbb J_\alpha$ are learned all at the same time (see Sec.~\ref{sec:during}) or sequentially (see Sec.~\ref{sec:after}). In both cases, the group-theoretic structure of the symmetries will be tested through suitable modifications in the loss function as discussed in Sec.~\ref{sec:loss_modification}.

\subsection{Inference while learning}
\label{sec:during}

From a theorist's point of view, the more principal approach is when all $N_g$ elements in a candidate set of symmetry generators $\{\mathbb G_\alpha\}$ ($\alpha=1,2,\ldots,N_g$) are learned in one go as in \cite{Forestano:2023fpj,Roman:2023ypv,Forestano:2023qcy}.  
The advantage of this one-shot approach is that throughout all stages of the training the collection of potential generators $\{\mathbb G\}$ is forced to constitute a closed algebra of the form 
\begin{equation}
\bigl[ \mathbb{G}_\alpha, \mathbb{G}_\beta\bigr] 
= \sum_{\gamma=1}^{N_g} a_{\alpha\beta}^{\gamma} \mathbb{G}_\gamma, 
\quad \forall \alpha, \beta \,= \,1,2,\ldots,N_g \, ,
\end{equation}
where $[.,.]$ is the Lie bracket and $a_{\alpha\beta}^{\gamma}$ are the respective structure constants. To this end, one includes the following closure term in the total loss function \cite{Forestano:2023fpj}  
\beq
L_\text{closure} \bigl(\{\mathbb G\}, a_{\alpha\beta}^{\gamma} \bigr) = 
\sum_{ \alpha=1}^{N_g-1} \sum_{\substack{\beta=\alpha+1 }}^{N_g} 
\Tr \left(\mathbb{C}_{\alpha\beta}  \cdot\mathbb{C}_{\alpha\beta}^\dagger\right),
\label{eq:LossClosure}
\eeq
where $\dagger$ denotes Hermitian conjugation. The violation of the closure condition is parametrized by 
\beq 
\mathbb{C}_{\alpha\beta} \bigl(\{\mathbb G\},a_{\alpha\beta}^{\gamma}\bigr) \equiv
\bigl[ \mathbb{G}_\alpha, \mathbb{G}_\beta\bigr] 
- \sum_{\gamma=1}^{N_g} a_{\alpha\beta}^{\gamma} \mathbb{G}_\gamma .
\label{eq:closuremismatch}
\eeq
With this approach, the subalgebra structure can be inferred concurrently with the deep learning process. 

\subsection{Inference as postprocessing}
\label{sec:after}

The alternative approach, where the generators are learned one at a time \cite{Moskalev2210.04345,Forestano:2023ijh}, is computationally faster, but then the inference of the subgroup structure must be performed as a postprocessing step. Suppose that the search for symmetries has resulted in a total of $N_{\mathfrak s}$ learned generators $\{\mathbb J_\alpha\}$ representing the full symmetry algebra $\mathfrak s$. Following \cite{Forestano:2023ijh}, we then rotate into a new basis $\{ \tilde {\mathbb J}\}$ as
\begin{equation}
\tilde {\mathbb J}_\alpha ({\mathbb O})= \sum_{\beta=1}^{N_{\mathfrak s}} {\mathbb O}_{\alpha\beta}\,{\mathbb J}_{\beta},
\label{eq:sparse_rotation}
\end{equation}
where ${\mathbb O}$ is a constant orthogonal matrix. The loss functions discussed below in Section~\ref{sec:loss_modification} will now be defined in terms of the rotated generators $\{\tilde {\mathbb J}\}$ and will therefore implicitly depend on the elements of the matrix ${\mathbb O}$, which are the parameters to be learned during the training. Since the original set of generators ${\mathbb J}_\alpha$ was already orthonormal, and we use a parametrization in which ${\mathbb O}$ is manifestly orthogonal, the new generators $\tilde {\mathbb J}_\alpha$ are guaranteed to be orthonormal as well, and there is no need to include orthogonality and normalization terms in the loss function. Furthermore, the invariance loss term is also not needed at this stage, since $\{\mathbb J\}$ are symmetry generators and any linear combination of them represents a symmetry as well. Therefore, the training for ${\mathbb O}$ is done with a loss function which only enforces closure (and optionally sparsity).

\subsection{Loss Function Modifications}
\label{sec:loss_modification}

\subsubsection{Finding the full symmetry algebra}
\label{sec:findingfull}

The first order of business is to find the full symmetry algebra $\mathfrak s$, i.e., the largest possible closed set of generators $\{\mathbb J\}$. As discussed in \cite{Forestano:2023fpj,Forestano:2023ijh}, this is a relatively straightforward exercise. In the one-shot approach of Section~\ref{sec:during}, the number of generators $N_g$ in the closure loss (\ref{eq:LossClosure}) is treated as a hyperparameter which is continually being incremented. Some values of $N_g$ result in successful training, others do not. Whenever a valid closed set of symmetry generators is found during this process, this guarantees the existence of a subalgebra $\mathfrak h$ with $N_{\mathfrak h} = N_g$ generators. The maximum obtained value, $N_{\mathfrak s} \equiv \max\{N_{\mathfrak h}\}$, of $N_{\mathfrak h}$, is the dimension of the full symmetry algebra $\mathfrak s$. In the sequential approach of Section~\ref{sec:after} the idea is very similar ---  one keeps trying to learn a new non-trivial symmetry generator which is orthogonal to the set found so far. When no such new generator can be found, the number of existing generators found by then is precisely $N_{\mathfrak s}$.

Note that both approaches (the one-shot learning from Section~\ref{sec:during} and the sequential learning from Section~\ref{sec:after}) result in the same outcome. The only difference is that in the former case, the loss functions are written in terms of the generator matrices $\mathbb G_\alpha$, whose components are the learnable parameters, while in the latter case, the loss functions are written in terms of the rotated generators $\tilde {\mathbb J}_\alpha(\mathbb O)$, and the learnable parameters are the components of the rotation matrix $\mathbb O$. In what follows we shall use the notation of Section~\ref{sec:during} and write our loss functions in terms of $\mathbb G_\alpha$. It should be understood that in the case of the sequential approach of Section~\ref{sec:after}, the same loss functions can be used, but with the replacement $\mathbb G_\alpha \to \tilde {\mathbb J}_\alpha(\mathbb O)$.

\subsubsection{Identifying the rank}
\label{sec:rank}

Next, we would like to find the rank of the thus found algebra $\mathfrak s$. The rank of a Lie algebra is the dimension of its Cartan subalgebra (the maximal {\em Abelian} subalgebra). In other words, we are looking for the largest subalgebra $\mathfrak h_c$ whose elements all commute with each other. In order to find $\mathfrak h_c$, we can repeat the previous exercise from Section~\ref{sec:findingfull}, only this time we set all structure constants $a_{\alpha\beta}^{\gamma}$ in (\ref{eq:closuremismatch}) to zero, resulting in
\begin{equation}
\mathbb{C}_{\alpha\beta} \bigl(\{\mathbb G\}\bigr) =
\bigl[ \mathbb{G}_\alpha, \mathbb{G}_\beta\bigr].
\label{eq:Cabelian}
\end{equation} 
Under those conditions, the maximal allowed value for the hyperparameter $N_g$ will give the dimension $N_{\mathfrak h_c}$ of the Cartan subalgebra.

\subsubsection{Testing the subalgebra structure}
\label{sec:subalgebra_structure}

\begin{table}[t!]
\centering
\scalebox{0.78}{
\renewcommand\arraystretch{1.5}
\begin{tabular}{||c||c|c|c|c||}
\cline{2-5}
\multicolumn{1}{c||}{} & \multicolumn{4}{c||}{Number of subalgebra factors $h$} \\ \hline
$N_{\mathfrak h}$ & 1 & 2 & 3 & 4 \\  \hline\hline
1 & $\mathfrak h_1^{(1)}$ & --- & --- & ---  \\  \hline
2 & $\mathfrak h_1^{(2)}$ & $\mathfrak h_1^{(1)} \oplus \mathfrak h_2^{(1)}$ & --- & --- \\ \hline
3 & $\mathfrak h_1^{(3)}$ & $\mathfrak h_1^{(2)} \oplus \mathfrak h_2^{(1)}$ & $\mathfrak h_1^{(1)}\oplus\mathfrak h_2^{(1)}\oplus \mathfrak h_3^{(1)}$ & --- \\  \hline
\multirow{2}{*}{4} & \multirow{2}{*}{$\mathfrak h_1^{(4)}$} & $\mathfrak h_1^{(3)}\oplus \mathfrak h_2^{(1)}$ & \multirow{2}{*}{$\mathfrak h_1^{(2)}\oplus\mathfrak h_2^{(1)}\oplus \mathfrak h_3^{(1)}$} &  \multirow{2}{*}{$\mathfrak h_1^{(1)}\oplus\mathfrak h_2^{(1)}\oplus \mathfrak h_3^{(1)}\oplus \mathfrak h_4^{(1)}$} \\ 
 &  & $\mathfrak h_1^{(2)}\oplus \mathfrak h_2^{(2)}$ &  & \\  \hline
$\vdots$ & \multicolumn{4}{c||}{$\vdots$} \\
\hline\hline
\end{tabular}
}
\caption{The setup for the subalgebra search discussed in Section~\ref{sec:subalgebra_structure}. For each possible integer value $N_{\mathfrak h}$ of the total number of generators in a subalgebra $\mathfrak h$, we consider all possible partitions into $h\le N_{\mathfrak h}$ distinct subgroups, each subgroup being a closed subalgebra $\mathfrak h_i^{(N_{\mathfrak h_i})}$ with $N_{\mathfrak h_i}$ generators.
\label{tab:setup}}
\end{table}

The procedure outlined in Section~\ref{sec:findingfull} already singles out the allowed values for the number of generators  $N_{\mathfrak h}$ in the allowed subalgebras of $\mathfrak s$. As depicted in Table~\ref{tab:setup}, we can now further probe the structure of these subalgebras $\mathfrak h$, by looking for factor decompositions. Specifically, we conjecture a partition of any sublgebra $\mathfrak h$ into a direct sum of $h$ subalgebras $\mathfrak h_1, \mathfrak h_2, \ldots, \mathfrak h_h$: 
\begin{equation}
\mathfrak h
\; = \; \mathfrak h_{1} \oplus \mathfrak h_{2} \oplus \cdots \oplus \mathfrak h_{h} \, . 
\label{eq:decomposition}   
\end{equation}
We shall label the number of generators in each subalgebra $\mathfrak h_i$ with $N_{\mathfrak h_i}$ (in Table~\ref{tab:setup} this value is listed as a parentheses-enclosed superscript). Therefore, the decomposition (\ref{eq:decomposition}) implies
\begin{equation}
N_{\mathfrak h} = N_{\mathfrak h_1}  + N_{\mathfrak h_2} + \ldots + N_{\mathfrak h_h}.
\end{equation}
In the special case when $\mathfrak h$ represents the full algebra, this procedure will give its decomposition, $\mathfrak s
= \mathfrak h_{1} \oplus \mathfrak h_{2} \oplus \cdots \oplus \mathfrak h_{h}$.
Note that often there are several inequivalent ways to partition $N_{\mathfrak h}$ generators into $h$ groups. Table~\ref{tab:setup} shows one such example already at $N_{\mathfrak h}=4$ and $h=2$: we can split the 4 generators into groups of $3+1$ or $2+2$. In our numerical experiments in the next two sections, we consider all possible such partitions.

The decomposition (\ref{eq:decomposition}) implies that i) generators belonging to two different subalgebras $\mathfrak h_i$ and $\mathfrak h_j$ with $i\ne j$ necessarily commute, and ii) that the Lie bracket of two generators belonging to the same subalgebra $\mathfrak h_i$ must close on the generators from that subalgebra. We can combine these two requirements together by modifying the closure loss~(\ref{eq:LossClosure}) as follows 
\begin{align}
L_\text{closure} \bigl(\{\mathbb G\}, a_{\alpha\beta}^{\gamma} \bigr) = \sum_{i=1}^{h}\sum_{j=1}^{h} \sum_{\alpha = 1}^{N_{\mathfrak h_i}} \sum_{\beta = 1}^{N_{\mathfrak h_j}}
\Tr \left(\mathbb{C}_{\alpha\beta}^{(ij)}  \cdot \left(\mathbb{C}^{(ij)}_{\alpha\beta}\right)^\dagger\right),
\label{eq:LossClosure2}
\end{align}
where
\beq 
\mathbb{C}_{\alpha \beta}^{(ij)} \equiv
\bigl[ \mathbb{G}_\alpha^{(i)}, \mathbb{G}_\beta^{(j)}\bigr] 
- \delta_{ij}\sum_{\gamma=1}^{N_{\mathfrak h_i}} \left(a_{i}\right)_{\alpha \beta}^{\gamma} \mathbb{G}_\gamma^{(i)} .
\label{eq:closuremismatch2}
\eeq
Here $a_i$ denotes the tensor of structure constants of the subalgebra $\mathfrak h_i$. The upper indices $(i)$ and $(j)$, while redundant, serve as useful reminders that the index $\alpha$ runs over the generators in $\mathfrak h_i$, while the index $\beta$ runs over the generators in $\mathfrak h_j$.

\section{\texorpdfstring{$U(4)$ subalgebra structure}{U(4) subalgebra structure}}
\label{sec:SU4example}

\begin{figure}[t]
\centering
\includegraphics[width=0.95\columnwidth]{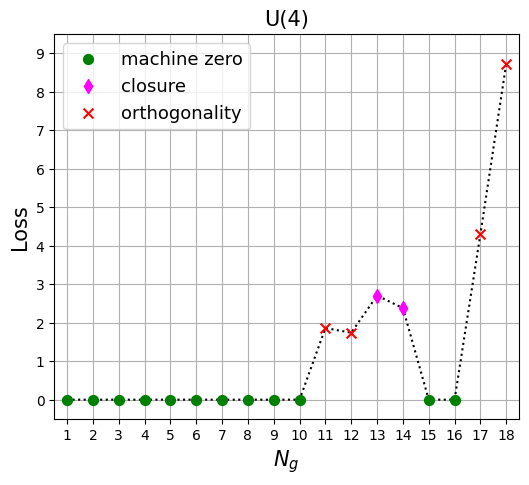}
\caption{The final value of the loss function as a function of the requested number of generators $N_g$ for the $U(4)$ example of Section~\ref{sec:SU4example}. The colored symbols identify the dominant contribution to the total loss: magenta diamonds for closure and red crosses for orthogonality. For the green circles the total loss is zero (to within machine precision). The learning rate was 0.001 and the training was done for 7,000 epochs. } \label{fig:loss}
\end{figure}

We now illustrate the techniques described in the previous section with a specific example, for which we chose the $U(4)$ unitary group (although not shown here, we also worked out the cases of $U(2)$, $U(3)$ and $U(5)$). While perhaps not as popular as its ``neighbors" $U(3)$ and $U(5)$, $U(4)$ has found applications in various areas of physics \cite{Akulov:1972zd,Tilma2002APO,PhysRevA.74.030304,PhysRevA.77.032332} and a complete account of its subalgebras is available \cite{Ivanov1985}. For our purposes, $U(4)$ strikes a nice balance between the relative simplicity of the widely used $U(2)$ and $U(3)$ groups, and the complexity of $U(5)$ and higher groups used in GUT model building. Furthermore, $U(4)$ already contains an interesting non-trivial subgroup, namely $SO(5) \sim Sp(4)$, which is described in detail in \ref{sec:so5}.

The $U(4)$ symmetry results from the following oracle defined on ${\mathbf x}\in \mathbb C^4$
\beq
\varphi_{U}(\mathbf{x}) 
\; \equiv \; |\mathbf{x}|^2 \; = \; \sum_{j=1}^4 \bigl(x^{(j)}\bigl)^\ast x^{(j)}, 
\quad 
x^{(j)}\in \mathbb C.
\label{oracle:U}
\eeq

Using the procedure from Section~\ref{sec:findingfull}, we vary the hyperparameter $N_g$ to find the allowed values for the number of generators $N_{\mathfrak h}$ in a subalgebra $\mathfrak h$. The result is shown in Figure~\ref{fig:loss}. We see that the largest possible number of generators in this case is $N_{\mathfrak s}=16$, which corresponds to the full algebra $\mathfrak s=u(4)$. Based on the low values of the loss, we conclude that there exist subalgebras with any $N_{\mathfrak h}$ from 1 to 10, and also $N_{\mathfrak h}=15$. Our main task now will be to decipher exactly what type of subalgebras are those.

\begin{figure}[t!]
    \centering
    \includegraphics[width=0.47\textwidth]{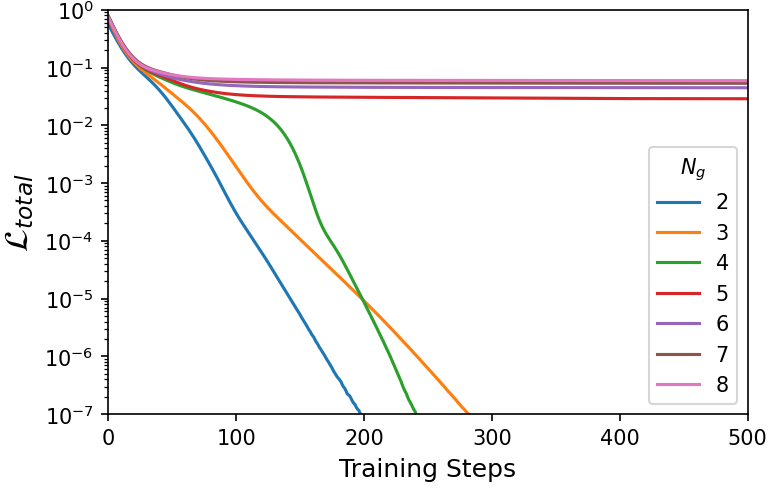}
    \caption{Finding the Cartan subalgebra of the $u(4)$ algebra. The evolution of the value of the total loss function with the Abelian closure condition (\ref{eq:Cabelian}) is plotted for different number of generators $N_g$ as shown in the legend. }
    \label{fig:algebrarankloss}
\end{figure}

Next we use the method of Section~\ref{sec:rank} to determine the rank of the so found algebra $\mathfrak s$. Figure~\ref{fig:algebrarankloss} depicts the evolution of the value of the total loss function ${\cal L}_{total}$ as a function of the training step, with the Abelian closure condition $\bigl[ \mathbb{G}_\alpha, \mathbb{G}_\beta\bigr] = 0$ imposed. In order to find the maximal Abelian algebra, we increment the value of the number of candidate generators $N_g$, as listed in the legend. We see that the training is successful and the loss is driven to zero for $N_g=2$, 3 and 4. However, as soon as $N_g$ hits 5 or higher, the loss remains large, indicating that there is no valid abelian subalgebra of that size. We therefore conclude that the rank of the 16-dimensional full algebra discovered in the previous step is 4, which is precisely the result expected from Lie group theory.

\definecolor{mylime}{RGB}{190, 240, 170}
\definecolor{myorange}{RGB}{255, 225, 150}

\begin{table}[t]
\centering
\scalebox{0.8}{
\renewcommand\arraystretch{1.4}
\begin{tabular}{||c||c|c|c|c||}
\cline{2-5}
\multicolumn{1}{c||}{} & \multicolumn{4}{c||}{Number of subalgebra factors $h$} \\ 
\hline
$N_{\mathfrak h}$ & 1 & 2 & 3 & 4 \\
\hline\hline
   1 & 
   \cellcolor{mylime}$u_1$ 
   & --- & --- & ---  \\  \hline
 2 & \cellcolor{mylime}{$u_1^2$} & \cellcolor{mylime}$u_1\oplus u_1$ & --- & --- \\ \hline
 \multirow{2}{*}{3} & \cellcolor{mylime}$u_1^3$ & \cellcolor{mylime}{$u_1^2\oplus u_1$} & \cellcolor{mylime}{$u_1\oplus u_1\oplus u_1$} & \multirow{2}{*}{---} \\  
   & \cellcolor{mylime}$su_2$ & \cellcolor{myorange} & \cellcolor{myorange} &  \\  \hline
\multirow{2}{*}{4} & \cellcolor{mylime}{$u_1^4$} & \cellcolor{mylime}$u_1^3\oplus u_1$ & \cellcolor{mylime}{{$u_1^2\oplus u_1\oplus u_1$}} & \cellcolor{mylime}{$u_1\oplus u_1\oplus u_1\oplus u_1$}  \\ 
  & \cellcolor{mylime}$u_2$  & \cellcolor{mylime}$su_2\oplus u_1$ & \cellcolor{myorange} & \cellcolor{myorange} \\ 
\hline
5 & \cellcolor{mylime}{$\rightarrow$} & \cellcolor{mylime}{$u_2\oplus u_1$} & \cellcolor{mylime}{$su_2\oplus u_1\oplus u_1$} &  \cellcolor{myorange}{}\\  \hline 
\multirow{2}{*}{6} & \cellcolor{mylime}{$\rightarrow$} & \cellcolor{mylime}{$\rightarrow$} & \cellcolor{mylime}{$u_2\oplus u_1\oplus u_1$} & \cellcolor{mylime}{$su_2\oplus u_1\oplus u_1\oplus u_1$}\\ 
  & \cellcolor{mylime}{$so_4$} & \cellcolor{mylime}{$su_2\oplus su_2$} & \cellcolor{myorange}{} & \cellcolor{myorange}{}\\  \hline 
 7 & \cellcolor{mylime}{$\rightarrow$}        & \cellcolor{mylime}{$u_2\oplus su_2$} & \cellcolor{mylime}{$su_2\oplus su_2\oplus u_1$} & \cellcolor{myorange}  \\  \hline
 \multirow{2}{*}{8} & \cellcolor{mylime}$\rightarrow$ & \cellcolor{mylime}{$u_2\oplus u_2$} & \cellcolor{mylime}{$u_2\oplus su_2\oplus u_1$} & \cellcolor{mylime}$su_2\oplus su_2\oplus u_1\oplus u_1$  \\  
& \cellcolor{mylime}$su_3$ & \cellcolor{myorange} & \cellcolor{myorange} & \cellcolor{myorange}  \\  
 \hline
9 & \cellcolor{mylime}$u_3$ & \cellcolor{mylime}{$su_3\oplus u_1$} & \cellcolor{myorange}{} & \cellcolor{myorange}  \\  \hline
\multirow{2}{*}{10} & \cellcolor{mylime}$\rightarrow$        & \cellcolor{mylime}{$u_3\oplus u_1$} & \cellcolor{mylime}{$su_3\oplus u_1\oplus u_1$} & \cellcolor{myorange}  \\ 
  & \cellcolor{mylime}$sp_4$        & \cellcolor{myorange} & \cellcolor{myorange} & \cellcolor{myorange}  \\ \hline
11 & \cellcolor{myorange}        & \cellcolor{myorange}{} & \cellcolor{myorange}{} & \cellcolor{myorange}  \\  \hline 
12 & \cellcolor{myorange}        & \cellcolor{myorange}{} & \cellcolor{myorange}{} & \cellcolor{myorange}  \\  \hline 
13 & \cellcolor{myorange}        & \cellcolor{myorange}{} & \cellcolor{myorange}{} & \cellcolor{myorange}  \\  \hline
14 & \cellcolor{myorange}        & \cellcolor{myorange}{} & \cellcolor{myorange}{} & \cellcolor{myorange}  \\  \hline
15 & \cellcolor{mylime}{$su_4$}        & \cellcolor{myorange}{} & \cellcolor{myorange}{} & \cellcolor{myorange}  \\  \hline
16 & \cellcolor{mylime}{$u_4$}       & \cellcolor{mylime}{$su_4\oplus u_1$} & \cellcolor{myorange}{} & \cellcolor{myorange}  \\    
 \hline\hline
\end{tabular}
}
\caption{The subalgebra decomposition results for the case of $u(4)$ presented in analogy to Table~\ref{tab:setup}. The viable partitions are only up to $h=4$, which is the rank of the full algebra. Green (yellow) boxes indicate the existence (the absence) of a valid decomposition. Results appearing on the same row are isomorphic to each other, while results with the same $N_{\mathfrak h}$, but on different rows represent different (non-isomorphic) subalgebras.
\label{tab:U4}}
\end{table}

Finally, we apply the technique of Section~\ref{sec:subalgebra_structure} to obtain the decompositions (\ref{eq:decomposition}) of the valid subalgebras found in Fig.~\ref{fig:loss}. The results are summarized in Table~\ref{tab:U4}, and depend on the value of the number of generators $N_{\mathfrak h}$ in the subalgebra, and the number of subalgebra groups $h$. For compactness, in Table~\ref{tab:U4} we use Slansky notation \cite{Slansky:1981yr}, where subscripts denote the dimensionality $n$, i.e., $u_4\equiv u(4)$, $su_4\equiv su(4)$, $sp_4\equiv sp(4)$, etc. Green (yellow) boxes in the table indicate the existence (the absence) of a valid decomposition. For example, we confirm the result from Fig.~\ref{fig:loss} that there are no closed subalgebras with $N_{\mathfrak h}=11$, 12, 13 or 14 generators. In the remaining cases, we do find valid subalgebras, which, as a rule, can themselves be further decomposed into factors (the only exceptions being the cases of $N_{\mathfrak h}=1$ and $N_{\mathfrak h}=15$). 

Note that sometimes there are two different non-isomorphic subalgebras for the same values of $N_{\mathfrak h}$ and $h$. Such cases appear as separate entries on different rows in the corresponding box in the table. For example, consider $N_{\mathfrak h}=6$ and $h=2$. There are three viable subalgebra decompositions into two groups: i) $3+3$, which is the case of $su_2\oplus su_2$; ii) $4+2$, which is the case of $u_2\oplus (u_1\oplus u_1)$; and iii) $5+1$, which is isomorphic to the previous case and is given by $(u_2\oplus u_1)\oplus u_1$.

\begin{figure*}[t!]
    \centering
    \includegraphics[width=1\linewidth]{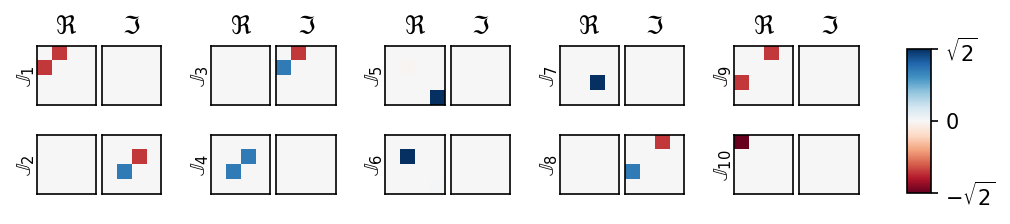}
    \caption{The learned generators for the $su_3\oplus u_1\oplus u_1$ case of $N_{\mathfrak h}=10$ in Table~\ref{tab:U4}. In this and all subsequent such figures, each learned generator matrix ${\mathbb J}_\alpha$ is represented by a pair of panels (one for the real and one for the imaginary parts). The values of the individual elements of the matrix are color-coded and can be read off the color bar. 
    }
    \label{U4_10_U3U1:mapping}
\end{figure*}

\begin{figure*}[t!]
    \centering
    \includegraphics[width=1\linewidth]{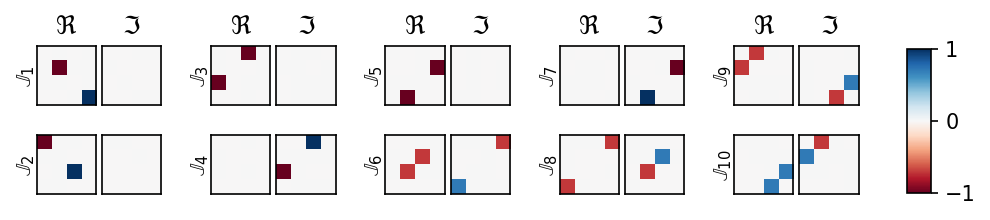}
    \caption{
    The same as Fig.~\ref{U4_10_U3U1:mapping}, but for the $sp_4\sim so_5$ case of $N_{\mathfrak h}=10$ in Table~\ref{tab:U4}.
    }
    \label{U4_10_SO5:mapping}
\end{figure*}

A particularly interesting case occurs for $N_{\mathfrak h}=10$ and $h=1$. We obtain two different subalgebras, the rank 4 subalgebra $su_3\oplus u_1 \oplus u_1$, which can be decomposed further into three factors, and the rank 2 subalgebra $sp_4 \sim so_5$, which is simple and cannot be decomposed further. Figures~\ref{U4_10_U3U1:mapping} and \ref{U4_10_SO5:mapping} show the learned sparse generators in those two cases, respectively. In Fig.~\ref{U4_10_U3U1:mapping}, the $su_3$ factor consists of 
$\mathbb J_1$,
$\mathbb J_2$,
$\mathbb J_3$,
$\mathbb J_4$,
$\mathbb J_8$,
$\mathbb J_9$,
and two traceless linear combinations of 
$\mathbb J_6$, $\mathbb J_7$, and $\mathbb J_{10}$.
In order to demonstrate that the learned generators of Fig.~\ref{U4_10_SO5:mapping} form an $so_5$ algebra, we can map them explicitly to the ten $so_5$ generators (\ref{eq:so5set}) discussed in~\ref{sec:so5} as follows
\begin{subequations}
\begin{align}
\mathbb J_{1} &= - \frac{1}{\sqrt{2}}
\left( L_{34} - L_{15}\right),\\
\mathbb J_{2} &= - \frac{1}{\sqrt{2}}
\left( L_{34} + L_{15} \right),\\
\mathbb J_{3} &= - \frac{1}{\sqrt{2}}
\left( L_{45} + L_{13} \right),\\
\mathbb J_{4} &= - \frac{1}{\sqrt{2}}
\left( L_{14} - L_{35}\right),\\
\mathbb J_{5} &= - \frac{1}{\sqrt{2}}
\left( L_{45} - L_{13} \right),\\
\mathbb J_{6} &= - \frac{1}{\sqrt{2}}
\left( L_{23} - L_{24}\right),\\
\mathbb J_{7} &= - \frac{1}{\sqrt{2}}
\left( L_{14} + L_{35}\right),\\
\mathbb J_{8} &= - \frac{1}{\sqrt{2}}
\left( L_{23} + L_{24}\right),\\
\mathbb J_{9} &= - \frac{1}{\sqrt{2}}
\left( L_{25} + L_{12} \right),\\
\mathbb J_{10} &= - \frac{1}{\sqrt{2}}
\left( L_{25} - L_{12} \right),
\end{align}
\end{subequations}
where $L_{ij}$ are given in the representation (\ref{ourSO5generators4x4}).

\section{Spontaneous symmetry breaking of non-abelian gauge symmetries}
\label{sec:models}

In this section we demonstrate the application of the symmetry finding and identification procedures from the previous sections to particle theory model building, using a couple of examples from the classic textbook \cite{Peskin:1995ev}.

\subsection{\texorpdfstring{$SU(3)$ Model}{SU(3) model}}
\label{sec:toysu3}

Following Chapter 20 in \cite{Peskin:1995ev}, consider an $SU(3)$ gauge theory with an adjoint scalar
\begin{equation}
\Phi = \sum_{a=1}^8 \phi_a \, t^a,
\end{equation}
where $t^a$ are the $3\times 3$ Hermitian matrices representing the generators in the adjoint representation of $SU(3)$. A gauge transformation $U$ acts on the Higgs field $\Phi$ as
\beq
\Phi \longrightarrow U \, \Phi\, U^\dagger.
\label{eq:gaugetransformation}
\eeq

The physics is contained in the potential $V$ of the theory, which in our case will play the role of the oracle $\varphi(\mathbf x)$. In turn, the role of the features $\mathbf x$ will be taken by the field components $\phi_a$. In order to generate spontaneous symmetry breaking, we can choose a potential
\beq
V(\phi) = \left[ \textrm{Tr} \left(\Phi^\dagger \Phi\right) - \frac{v^2}{2}\right]^2
\label{eq:potential}
\eeq
with a non-vanishing parameter $v$. At the minimum of this potential, $\Phi$ has a non-zero vacuum expectation value (vev), $\Phi_0\equiv \langle\Phi\rangle$, and breaks the $SU(3)$ symmetry spontaneously. Expanding around the vev as
\beq
\Phi \equiv \Phi_0 + \eta,
\label{eq:vevexpansion}
\eeq
the potential can be rewritten in terms of the physical degrees of freedom $\eta$ as
\beq
V(\eta) = \left[ \textrm{Tr} \left(\eta^\dagger \eta\right) 
+ \textrm{Tr}\left(\Phi_0 \eta\right)
+ \textrm{Tr}\left( \eta^\dagger \Phi_0\right)\right]^2.
\label{eq:etaoracle}
\eeq

In order to derive the symmetry of this model, we create a dataset by sampling the 8-dimensional complex vector $(\phi_1,\phi_2,\ldots,\phi_{8})$, then forming $\eta$ and looking for linearized transformations (\ref{eq:gaugetransformation}) of the form
\beq
\eta \longrightarrow \eta + i\,\varepsilon\, \mathbb{G}\,\eta - i\,\varepsilon\, \eta\, \mathbb{G}^\dagger
\label{eq:linearized_gauge_transformation}
\eeq
where the generator $\mathbb{G}$ is a $3\times 3$ complex Hermitian matrix.

\begin{figure}[t]
    \centering
    \includegraphics[width=1\linewidth]{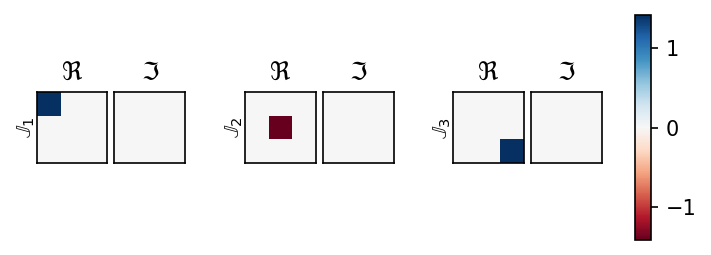}
    \caption{The symmetry generators after breaking $SU(3)$ with the adjoint vev (\ref{eq:SU3vev110}). }
    \label{fig:SU3_lambda3}
\end{figure}

Depending on the orientation of the vacuum-expectation value of the Higgs field, different symmetry breaking patterns may emerge. For example, if 
\begin{equation}
\Phi_0 \; = \; 
\frac{v}{2}\  \textrm{diag} \left( 1 , -1 , 0 \right),
\label{eq:SU3vev110}
\end{equation}
the $SU(3)$ gauge symmetry is broken down to $U(1)\times U(1)$. This situation is depicted in Fig.~\ref{fig:SU3_lambda3}, which shows the learned symmetry generators in that case.  The three diagonal generators shown in the figure can be used to form two traceless linear combinations which correspond to the two independent $U(1)$'s.

\begin{figure}[t]
    \centering
    \includegraphics[width=1\linewidth]{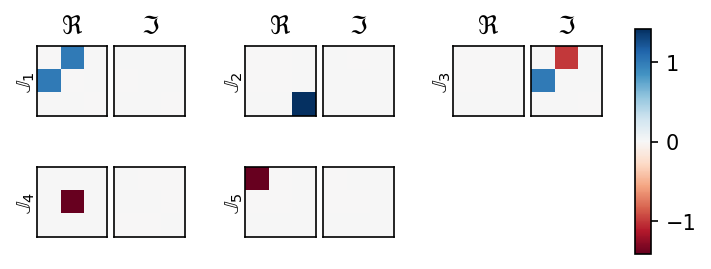}
    \caption{The symmetry generators after breaking $SU(3)$ with the adjoint vev (\ref{eq:SU3vev112}). }
    \label{fig:SU3_lambda8}
\end{figure}

Another possible choice for a symmetry breaking vacuum is 
\begin{equation}
\Phi_0 \; = \; 
\frac{v}{2\sqrt{3}}\  \textrm{diag} \left( 1 , 1 , -2 \right).
\label{eq:SU3vev112}
\end{equation}
As shown in Figure~\ref{fig:SU3_lambda8}, this results in the residual symmetry pattern $SU(2)\times U(1)$: $\mathbb J_1$, $\mathbb J_3$ and the antisymmetric combination of $\mathbb J_4$ and $\mathbb J_5$ combine to form the $SU(2)$ factor, while $2\mathbb J_2+\mathbb J_4+\mathbb J_5$ is the remaining $U(1)$ factor.

\subsection{\texorpdfstring{$SU(5)$ GUT Model}{SU(5) GUT model}}
\label{sec:SU5}

The previous example can be generalized to larger groups, and in particular $SU(5)$ grand unification (see Problem 20.1 in \cite{Peskin:1995ev}). The analysis from Section~\ref{sec:toysu3} goes through largely intact, the only difference being that the scalar field $\Phi$ is now expanded in terms of the 24 generators $T^a$ of the adjoint representation of $SU(5)$ \cite{raby_2021}
\beq
\Phi = \sum_{a=1}^{24} \phi_a\, T^a.
\eeq 
Using the same potential (\ref{eq:potential}) and expanding as in (\ref{eq:vevexpansion}), we again obtain the oracle (\ref{eq:etaoracle}), which now represents a map $\mathbb{C}^{24} \rightarrow{\mathbb{R}}$.

\begin{figure}[t!]
    \centering
    \includegraphics[width=1\linewidth]{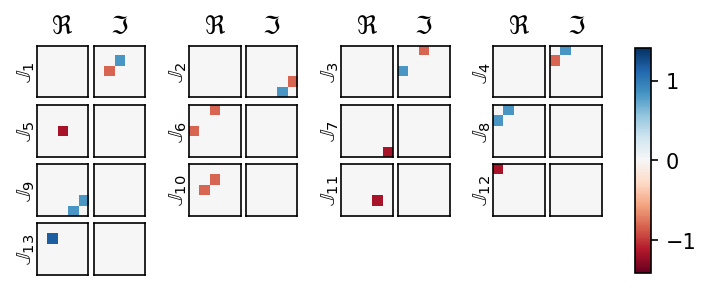}
    \caption{The symmetry generators after breaking $SU(5)$ with the adjoint vev (\ref{eq:T24vev}).}
    \label{fig:SU5_T24}
\end{figure}

The $SU(5)$ symmetry is spontaneously broken by a non-vanishing vev  for $\Phi$.
If the vev happens to be along the diagonal $T^{24}$ generator, i.e.,
\beq
\langle \Phi \rangle = v\, T^{24} 
\equiv v\, \sqrt{\frac{3}{5}}\ 
\textrm{diag} \left( -\frac{1}{3}, -\frac{1}{3}, -\frac{1}{3}, \frac{1}{2}, \frac{1}{2}\right),
\label{eq:T24vev}
\eeq
then the remaining symmetry is precisely that of the Standard Model, $SU(3)\times SU(2)\times U(1)$. 
In order to derive the residual symmetry in this case, we sample the 24-dimensional complex vector $(\phi_1,\phi_2,\ldots,\phi_{24})$, then form $\eta_a$ and look for transformations of the type (\ref{eq:linearized_gauge_transformation})
with $5\times 5$ complex Hermitian matrices $\mathbb{G}$. The result is shown in Fig.~\ref{fig:SU5_T24} and indeed corresponds to the SM gauge symmetry. For example, the $SU(3)$ of color is generated by 
$\mathbb J_1$,
$\mathbb J_3$,
$\mathbb J_4$,
$\mathbb J_6$,
$\mathbb J_8$,
$\mathbb J_{10}$,
$\mathbb J_{12}+\mathbb J_{13}$, and
$2 \mathbb J_5 - \mathbb J_{12}+\mathbb J_{13}$;
the weak $SU(2)$ is generated by 
$\mathbb J_2$,
$\mathbb J_9$ and 
$\mathbb J_{7}-\mathbb J_{11}$,
and the traceless $U(1)$ factor is
$2\mathbb J_5 - 3\mathbb J_{7} -3\mathbb J_{11} + 2\mathbb J_{12}-2\mathbb J_{13}$.

\section{Conclusions}
\label{sec:conclusions}

The research presented in this letter is the natural extension of the program started in \cite{Forestano:2023fpj,Roman:2023ypv,Forestano:2023qcy,Forestano:2023ijh} of using machine learning techniques to find symmetries in data or theory. We showed how the group-theoretic structure of the learned symmetry generators can be identified either during the learning process, or as a post-processing procedure. The approach was outlined in Section~\ref{sec:method} and subsequently illustrated with examples from group theory (in Section~\ref{sec:SU4example}) and from particle physics (in Section~\ref{sec:models}). The obtained insights into the learned symmetries offer clarity and explainability to the machine learning methodology.

{\bf Acknowledgements.}
We thank S.~Gleyzer, R.~Houtz, K.~Kong, S.~Mrenna, H.~Prosper and P.~Shyamsundar for useful discussions. We thank P.~Ramond for group theory insights and inspiration.  This work is supported in part by the U.S.~Department of Energy award number DE-SC0022148.

\appendix

\section{The \texorpdfstring{$SO(5)$}{SO(4)} subgroup of  \texorpdfstring{$SU(4)$}{SU(4)}.}
\label{sec:so5}

\begin{figure}[t]
    \centering
    \includegraphics[width=0.7\linewidth]{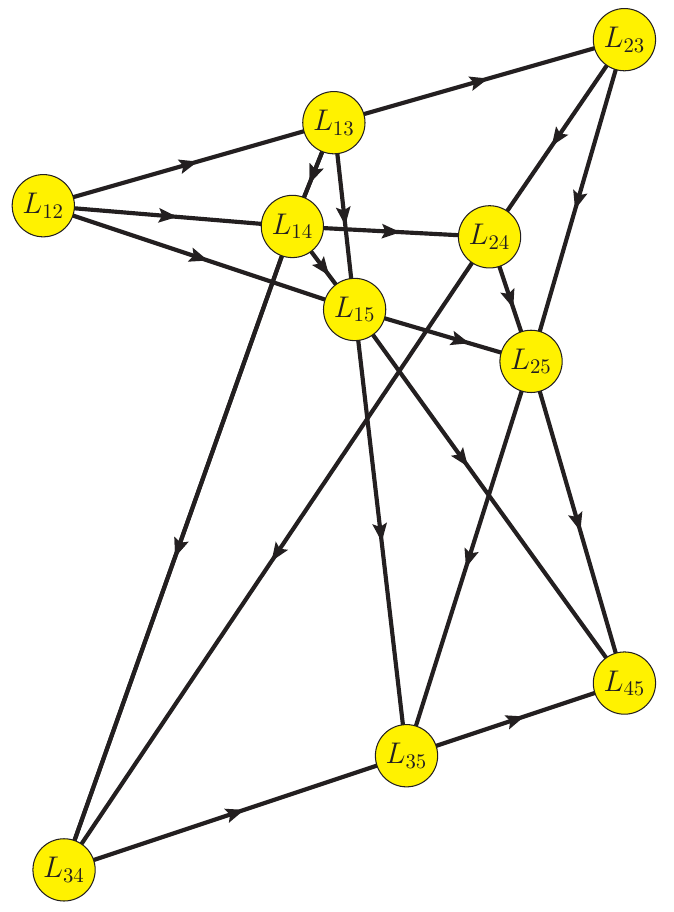}
    \caption{A pictorial representation of the $so(5)$ algebra (\ref{eq:SO5algebra}) in terms of a Desargues configuration. }
    \label{fig:desargues}
\end{figure}

The $so(5)$ algebra is given by
\begin{equation}
[L_{mn}, L_{pq}]= i \left(
 \delta_{mp} L_{nq}  
+\delta_{nq} L_{mp}  
-\delta_{mq} L_{np}  
-\delta_{np} L_{mq}  
\right).
\label{eq:SO5algebra}
\end{equation}
Here each of the ten elements of the algebra, $L_{mn}$, is labelled by an antisymmetric pair of indices $mn$, where  $m,n\in \{1,2,3,4,5\}$ and $m\ne n$. Since $L_{mn}=-L_{nm}$, for concreteness and without loss of generality we can take the defining set of 10 independent generators of $SO(5)$ to be those with $m<n$, i.e.
\begin{equation}
\big\{ L_{12},\, L_{13},\, L_{14},\, L_{15},\, L_{23},\, L_{24},\, L_{25},\, L_{34},\, L_{35},\, L_{45}\big\}.
\label{eq:so5set}
\end{equation}
Interestingly, the $so(5)$ algebra (\ref{eq:SO5algebra}) can be neatly summarized with a Desargues configuration as illustrated in Figure~\ref{fig:desargues} \cite{PhysRevA.61.032301,PhysRevA.79.042323}. The 10 generators (\ref{eq:so5set}) are used as the 10 points of the configuration and the directed lines encode {\em all of} the commutation relations (\ref{eq:SO5algebra}) --- the commutator of any two generators on the same line equals $i$ times the third generator on the line, with the sign being $+$ or $-$, depending on whether we are following or going against the arrow. Any two generators which are not connected by a line in the diagram, commute.

$SO(5)$ is the group of rotations in ${\mathbb R}^5$. Therefore, it has a representation in terms of $5\times 5$ orthogonal matrices whose generators are given by
\begin{equation}
\left(L_{mn}\right)_{ij}=-i\left(\delta_{m i }\delta_{nj}-\delta_{m j }\delta_{n i}\right),
\label{eq:SO5generators5x5}
\end{equation}
where $i$ and $j$ are the matrix indices ($i,j\in \{1,2,3,4,5\}$) indicating the plane of rotation in ${\mathbb R}^5$. 
Explicitly,
\begin{equation}
L_{12} =
\begin{pmatrix}
0 & -i & 0 & 0 & 0\\
i & 0 & 0 & 0 & 0\\
0 & 0 & 0 & 0 & 0\\
0 & 0 & 0 & 0 & 0 \\
0 & 0 & 0 & 0 & 0 
\end{pmatrix},
\label{eq:5x5matrices}
\end{equation}
and so on for the remaining generators in (\ref{eq:so5set}).

According to Table~\ref{tab:U4}, $so(5)$ is a subalgebra of $u(4)$. Therefore, in addition to the $5\times 5$ representation (\ref{eq:5x5matrices}), the $so(5)$  algebra must also have a representation in terms of $4\times 4$ complex unitary matrices. Indeed, one such representation is given by \cite{PhysRevA.61.032301,PhysRevA.79.042323}
\begin{subequations}
\begin{align}
L_{13}  &= \frac{1}{\sqrt{2}} \sigma_3 \otimes \sigma_1, 
&L_{34} &= \frac{1}{\sqrt{2}} I_2 \otimes \sigma_3, 
\\
L_{14} &= \frac{1}{\sqrt{2}} \sigma_3 \otimes \sigma_2, 
&L_{35} &= -\frac{1}{\sqrt{2}}I_2 \otimes \sigma_2, 
\\
L_{15} &= \frac{1}{\sqrt{2}} \sigma_3 \otimes \sigma_3, 
&L_{45} &= \frac{1}{\sqrt{2}} I_2 \otimes \sigma_1, 
\\
L_{23} &= \frac{1}{\sqrt{2}} \sigma_1 \otimes \sigma_1, 
&L_{12} &= \frac{1}{2} \sigma_2 \otimes I_2,  \\
L_{24} &= \frac{1}{\sqrt{2}}\sigma_1 \otimes \sigma_2,  
& & \\
L_{25} &= \frac{1}{\sqrt{2}} \sigma_1 \otimes \sigma_3, 
&  & 
\end{align}
\label{eq:RausGenerators}
\end{subequations}
Here $I_2$ is the $2\times2$ unit matrix, $\sigma_i$ are the Pauli matrices 
\begin{equation}
\sigma_1 = 
\begin{pmatrix}
0 & 1 \\
1 & 0
\end{pmatrix}, \quad
\sigma_2 = 
\begin{pmatrix}
0 & -i \\
i & 0
\end{pmatrix}, \quad
\sigma_3 = 
\begin{pmatrix}
1 & 0 \\
0 & -1
\end{pmatrix}
\nonumber
\end{equation}
and $\otimes$ stands for tensor product, e.g.,
\begin{equation}
L_{12} 
= \frac{1}{2} \sigma_2 \otimes I_2 
= \frac{1}{2}
\begin{pmatrix}
0 & 0 & -i & 0 \\
0 & 0 & 0 & -i \\
i & 0 & 0 & 0 \\
0 & i & 0 & 0 
\end{pmatrix}
\nonumber
\end{equation}

The Desargues diagram of Figure~\ref{fig:desargues} helps understand the $3+3+3+1$ pattern of the representation (\ref{eq:RausGenerators}). First, one chooses a Pauli matrix, in this case $\sigma_2$, and associates it with the center of perspectivity $L_{12}$ of the configuration via the product $\sigma_2\otimes I_2$. The other two Pauli matrices, $\sigma_3$ and $\sigma_1$, are respectively used to form the two triangles in central perspective via the products $\sigma_3\otimes\sigma_i$ and $\sigma_1\otimes\sigma_i$, $i=1,2,3$. Finally, the axis of perspectivity is formed by the Pauli matrices themselves (more precisely, by the products $I_2\otimes \sigma_i$, $i=1,2,3$)

However, there are other equivalent $4\times 4$ representations of the $so(5)$ algebra which can be built out of $I_2$ and the Pauli matrices following the same pattern. For example, the particular representation which was obtained in Fig.~\ref{U4_10_SO5:mapping}, corresponds to associating the sum $\sigma_1+\sigma_2$ with the center of perspectivity, and proceeding to build the rest of the representation as described above. Concretely,
\begin{subequations}
\begingroup
\addtolength{\jot}{1em}
\begin{align}
L_{12} & = \frac{1}{2} I_2  \otimes (\sigma_1+\sigma_2), \\ 
L_{13} & = \frac{1}{\sqrt{2}} \sigma_1 \otimes \sigma_3, \\
L_{14} & = \frac{1}{\sqrt{2}} \sigma_2 \otimes \sigma_3, \\
L_{15} & = \frac{1}{\sqrt{2}} \sigma_3 \otimes \sigma_3, \\
L_{23} & = \frac{1}{2} \sigma_1 \otimes (\sigma_1-\sigma_2), \\
L_{24} & = \frac{1}{2} \sigma_2 \otimes (\sigma_1-\sigma_2), \\
L_{25} & = \frac{1}{2} \sigma_3 \otimes (\sigma_1-\sigma_2), \\
L_{34} & = \frac{1}{\sqrt{2}} \sigma_3 \otimes I_2,  \\
L_{35} & = -\frac{1}{\sqrt{2}} \sigma_2 \otimes I_2, \\
L_{45} & = \frac{1}{\sqrt{2}} \sigma_1 \otimes I_2.
\end{align}
\endgroup
\label{ourSO5generators4x4}
\end{subequations}
The fact that the ten matrices (\ref{eq:RausGenerators}) or (\ref{ourSO5generators4x4}) form an $so(5)$ algebra may not be immediately obvious, but can be easily verified by substituting into the $so(5)$ commutation relations (\ref{eq:SO5algebra}) and checking that all 90 of them are identically satisfied (with a structure constant $\sqrt{2}$ instead of $1$).

 \bibliographystyle{elsarticle-num} 
 \bibliography{references}





\end{document}